Carlos Alberto Barrera-Diaz[a]*, Amir Nourmohammdi[a], Henrik Smedberg[a], Tehseen Aslam[a] and Amos H.C. Ng[a]

[a]*Intelligent Production Systems Division, University of Skövde, Skövde, Sweden;*

Corresponding author: Carlos Alberto Barrera Diaz (carlos.alberto.barrera.diaz@his.se)

# An enhanced simulation-based multi-objective optimization approach with knowledge discovery for reconfigurable manufacturing systems

In today's uncertain and competitive market, where enterprises are subjected to increasingly shortened product life-cycles and frequent volume changes, reconfigurable manufacturing systems (RMS) applications play a significant role in the manufacturing industry's success. Despite the advantages offered by RMS, achieving a high-efficiency degree constitutes a challenging task for stakeholders and decision-makers when they face the trade-off decisions inherent in these complex systems. This study addresses work tasks and resource allocations to workstations together with buffer capacity allocation in RMS. The aim is to simultaneously maximize throughput and minimize total buffer capacity under fluctuating production volumes and capacity changes while considering the stochastic behavior of the system. An enhanced simulation-based multi-objective optimization (SMO) approach with customized simulation and optimization components is proposed to address the abovementioned challenges. Apart from presenting the optimal solutions subject to volume and capacity changes, the proposed approach support decision-makers with discovered knowledge to further understand the RMS design. In particular, this study presents a problem-specific customized SMO combined with a novel flexible pattern mining method for optimizing RMS and conducting post-optimal analyzes. To this extent, this study demonstrates the benefits of applying SMO and knowledge discovery methods for fast decision-support and production planning of RMS.

## 1 Introduction

In today's volatile market, manufacturing enterprises are often challenged by ever-shortening product life-cycles together with unpredictable demands and fluctuating production volumes (Diaz, Aslam, and Ng 2021). Therefore, the ability of a manufacturing system to react and adjust its capacities and equipment to cope with the suddenly arising challenging variations encompasses a crucial consideration for production organizations (J. Dou et al. 2020; Koren, Wang, and Gu 2017). To cope with challenges generated by a dynamic market wherein variations in the demand need to be addressed, the concept of Reconfigurable Manufacturing Systems (RMS) was introduced by Koren et al. (1999). RMS are responsive production systems that, through reconfigurations, can efficiently add, remove, or relocate resources and equipment in response to market shifts (Diaz et al. 2020; Koren, Gu, and Guo 2018a). In a nutshell, RMS are essential to cost-efficiently achieve high flexibility and provide better scalability and responsiveness than traditional production systems. Many recent studies have pointed out that RMS research is a mainstream topic and a major drive toward the future of the manufacturing industry because companies need to evade the significant investment loss caused by non-utilized equipment when facing dynamic market demands (Fan et al. 2022; Bortolini, Galizia, and Mora 2018; Yelles-Chaouche et al. 2020).

An RMS consists of several workstations (WSs) that contain several parallels and identical resources (Koren, Gu, and Guo 2018b). When a single-product type is produced the RMS is classified as a single-part flow line (SPFL), and when several product types are produced, it is classified as a multi-part flow line (MPFL) (J. Dou, Dai, and Meng

2010; Goyal, Jain, and Jain 2012). In the automotive industry, where several product types are produced in the same system, the adoption of MPFL is increasingly common (J. Dou et al. 2020). Studies focused on the resource and task assignment of MPFL-RMS are scarce and overlook the buffer allocation problem.

As a consequence of the success in modelling and analyzing production systems, simulation methods have been widely employed within the manufacturing industry (Pehrsson et al. 2015). In the era of digitalization, simulation models are essential to better understand and assess the complex nature inherent in the dynamics scenarios found in manufacturing systems (Mourtzis 2020). Simulation modelling, particularly Discrete-Event Simulation (DES), is considered an effective tool for handling the uncertain and changing scenarios of manufacturing systems (Petroodi et al. 2019; Juan et al. 2015). Additionally, optimization techniques have been used to solve the NP-hard combinatorial problems found in RMS. However, despite the success shown by simulation and optimization techniques, researchers have shown their shortcomings when employed separately. As the complexity of the system and the decision variables increase, the use of simulation might become impractical (Niño-Pérez et al. 2018; Xu et al. 2016). On the other hand, most RMS optimization studies simplified the problem by disregarding the variability and stochasticity of the systems. Against this backdrop, simulation-based optimization (SBO) emerged as a combination of simulation and optimization, providing solutions to bigger-scale problems. SBO investigates an extensive decision space searching for an optimal or near-optimal input variables combination (Barrera Diaz et al. 2021). Simulation-based multi-objective optimization (SMO) is employed when several conflicting objectives exist. Although prior studies have applied optimization to deal with the RMS configuration challenges, the use of SMO is very sporadic. Researchers identify opportunities in using SMO to tackle the NP-hard difficulties of RMS (Bortolini, Galizia, and Mora 2018; A. Bensmaine, Dahane, and Benyoucef 2013). These combinatorial problems have been approached using exact and heuristic methods, whereas meta-heuristics methods, in particular, Genetic Algorithms (GAs) for single-objectives and Non-dominated Sorting Genetic Algorithm II (NSGA-II) (Deb et al. 2002) for multi-objectives problems have proven to achieve better performance in nearing the optimal solutions in the RMS field (Michalos, Makris, and Mourtzis 2012; Renzi et al. 2014; Delorme, Malyutin, and Dolgui 2016).

For RMS where many scenarios are involved, SMO generates complex and large datasets that are difficult to analyze. Knowledge-Driven Optimization (KDO) is a recent research area where data mining methods are used on the resulting SMO datasets to expose underlying knowledge regarding what constitutes the preferred solutions in accordance with the generated Pareto-optimal front. Decision-makers can benefit from using data mining methods to extract the patterns that support a better understanding of the problem under different circumstances (e.g., production volumes). (Bandaru, Ng, and Deb 2017a)

To this extent and because of the outstanding performance of the NSGA-II in facing multi-objective problems, this study contributes to the RMS research domain as follows:

(1) For MPFL-RMS, a customized NSGA-II is proposed with a problem-specific encoding and decoding strategy to optimize the system configuration subjected to scalable capacities and fluctuating production volumes by addressing the tasks assignment to WSs and the buffers allocation problem for maximum throughput (THP) and minimum total buffer capacity (TBC), simultaneously.

(2) To overcome inaccurate results and cope with RMS's dynamic and stochastic behavior (e.g., resources failure, variability of task times, and inter-station buffers) while simultaneously dealing with multiple objectives, the customized NSGA-II is incorporated with DES to render an SMO approach that takes the dynamic nature of RMS into account.

(3) This study enhanced the proposed SMO with a data mining methodology for post-optimal analysis on multi-dimensional and multi-objective optimization datasets by employing a novel flexible pattern mining algorithm in an industrial R&D RMS application. Knowledge will be extracted and represented in rules discovering the underlying patterns that constitute the preferred solutions for a scalable RMS under different production volumes.

The remaining of the article is structured as follows. Section 2 provides a frame of reference for the RMS challenges, the related work, and the understanding of SMO combined with multi-criteria decision-making and knowledge discovery. Section 3 mathematically formulates the RMS required information. The proposed approach and the tailor-made problem-specific procedures are presented in section 4. Section 5 introduces the case application, and section 6 shows the results and the discovered knowledge. Finally, the conclusions are presented in section 7.

## 2 Frame of reference

### 2.1 Reconfigurable Manufacturing Systems challenges and related work

According to Koren et al. (2018b), the three main challenging areas that RMS needs to address are the configuration of the system, the process planning, and the components of the system. The system configuration involves the physical arrangement of resources (e.g., equipment, machines, operators). The arrangement of the resources impact aspects such as the RMS's scalability, productivity, and functionality (Koren, Hu, and Weber 1998; Diaz et al. 2020). Most previous research is focused on the assignment of machines to WSs. The process planning area addresses the work tasks balancing and assignment throughout the system, affecting the reconfiguration efficiency to cope with production changes (e.g., scalable capacities and volume changes) (H. A. ElMaraghy 2007; Koren 2013). Research within process planning gravitates around the work task assignment optimization. Lastly, the components of the system deal with the required type and amount of resources, such as machines, operators, buffers, material handling, etc., to achieve the desired capacity (Koren, Gu, and Guo 2018a). This area is crucial for scaling the RMS and most of the existing research merely focuses on optimizing the number of resources (Diaz, Aslam, and Ng 2021). Accordingly, for an RMS to accommodate changes in production during its lifecycle, reconfigurations in these areas are required. Generally, prior research targets one or more of the areas mentioned above by reallocating, adding, removing resources and rebalancing the tasks in the system (Wang and Koren 2012; Koren, Wang, and Gu 2017). However, although these areas are studied, they are rarely addressed simultaneously. Some of the most relevant studies are reviewed below.

### 2.1.1 Single-part flow lines

Concerning SPFL, Shabaka and Elmaraghy (2008) presented a GA-based method for the process plan generation of RMS with cost as objective. In studies such as Dou et al. (2009) and (2011), the authors presented two GA approaches for optimizing the RMS configuration with the capital cost as an objective. A single-objective GA approach aiming at either maximizing THP or minimizing the number of machines employed in the system was presented in (Wang and Koren 2012). This study reconfigures a system without buffers, and tasks are rebalanced to meet specific production capacities. In a subsequent study, the authors presented five principles to design scalable RMS and extended their previous approach to include three cases in which the inter-station buffers had the same constant capacity (Koren, Wang, and Gu 2017). Moghaddam et al. (2018) introduced an integer linear programming model to select the optimal configuration based on cost. Deif & ElMaraghy (2007) also present a GA optimization study where cost is utilized as an objective, where the authors investigated managing the capacity scalability in RMS. Borisovsky et al. (2013) and Makssoud et al. (2013) presented two different single objective approaches, a GA approach and a mixed integer linear programming approach, which were utilized to find the best task allocation to minimize the capital cost and the number of machines respectively.

Contrary to the single objective studies, Goyal et al. (2012) presented a MOO NSGA-II-based approach for obtaining the optimal configuration regarding convertibility, cost, and resource utilization. In a subsequent study, Goyal & Jain (2016) extended their previous research by proposing a particle swarm optimization, which searched for an optimal set of SPFL configurations with the same optimization objectives as before: cost, resource utilization, and convertibility. A MOO approach for finding the optimal RMS configuration, in this case, simulation-based, was proposed by Diaz et al. (2020), wherein NSGA-II was employed for production rate, buffer capacity, and lead time optimization. Targeting the process plan area, Khezri et al. (2020) optimized the cost, production time, and sustainability using three different approaches, a *posteriori* augmented ε-constraint, and two evolutionary approaches, NSGA-II, and strength Pareto evolutionary algorithm. Touzout and Benyoucef (2018, 2019) employed exact and evolutionary methods to target the process plan area by optimizing the cost, time, and energy consumption during the machines' utilization.

### 2.1.2 Multi-part flow lines

In the context of MPFL, a cost-oriented study considering the availability of the machines was formulated by Youssef & ElMaraghy (2008) to address the RMS configuration problem using GA and Tabu search. The RMS configuration problem was approached again by J. Dou et al. (2010) using GA. A mathematical approach to minimize the RMS configuration design cost was developed by Saxena & Jain (2012). The cost was again the objective in an integer linear programming approach to find the optimal configuration design in a hypothetical part family proposed by Moghaddam et al. (2020). The RMS resource selection was approached by Bensmaine et al. (2013) using s simulation-based NSGA-II with completion time as the objective. Bensmaine et al. (2014) proposed a new heuristic approach focusing on the process plan of a MPFL with the makespan as the objective.

From a multi-objective optimization (MOO) perspective, Musharavati and Hamouda (2012) employed a simulated annealing algorithm to address the process plan generation and optimize cost and THP. Chaube et al. (2012) targeted the process plan area again using the ordinary NSGA-II with cost and time as objectives. Dou et al. (2016) targeted the flow line design of a MPFL using NSGA-II with cost and tardiness as objectives. The same objectives were optimized by Dou et al. (2020), introducing, this time, a particle swarm optimization approach.

## 2.2 Simulation-Based Multi-Objective Optimization and Multi-Criteria Decision Making

Multiple conflicting objectives offer many challenges for decision-makers in practical MOO scenarios. Not only do several objectives have to be optimized simultaneously to find a representative set of the Pareto-optimal front of solutions, but decision-makers need also to select a final trade-off solution to be implemented. Neither one is a trivial task. A decision-maker may have certain preferences about the solutions to a MOO problem (Miettinen, Hakanen, and Podkopaev 2016). When these preferences are known ahead of the optimization process, the decision-maker may employ *a priori* methods to focus the search on certain preferred regions. When preferences are unknown beforehand, *a posteriori* methods are used to find a representation of the entire Pareto-optimal front before the decision-maker begins to analyze the solutions and find a preferred region. Assuming no preference, multi-objective evolutionary algorithms (MOEAs) are a proficient tool for finding solutions that both converge close to the true Pareto-optimal front while also having a good spread over the front. For the decision-maker to perform an adequate *a posteriori* analysis of the solutions, a MOEA needs to live up to both requirements of convergence and diversity on the Pareto-optimal front (Deb 2014). SMO is the process of combining MOO and simulation. Combining these two techniques brings advantages to both fields (Jian and Henderson 2016). The general representation of an SMO problem is defined by several conflicting objectives included in the objective functions and possibly subjected to several equality and inequality constraints. The use of simulation allows the decision-maker to optimize a real-world system with much higher fidelity without the need to simplify the MOOP problem and lose intricate details and relationships that might exist in the system. Additionally, using MOEAs to solve SMO problems will discover and explore many more solutions than manual processes (Carson and Maria 1997).

## 2.3 Knowledge Discovery and Flexible Pattern Mining

Once the optimization process has ended, the decision-maker faces many trade-off solutions to analyze. Much of the literature focuses solely on analyzing the objective space while disregarding the role the decision space plays in a decision-maker's preferred solutions. It can be argued that more knowledge about both the objective space and the decision space and how they relate, generated with data mining and machine learning methods, will lead to more informed decision-making (Bandaru, Ng, and Deb 2017a).

Flexible Pattern Mining (FPM) (Bandaru, Ng, and Deb 2017b) is a recent rule mining method developed explicitly for knowledge discovery in MOO. FPM is based on the Apriori algorithm (Agrawal and Srikant 1994) and generates decision rules that describe a decision-maker's preferences regarding a *selected* and *unselected* set of solutions. Typically, the decision-maker chooses the *selected* set as the preferred non-dominated solutions and the *unselected* set as the remaining solutions found in the search. The FPM procedure then finds rules that discriminate the *selected* set from the *unselected* set, on the forms $x < c_1$, $x > c_2$, and $x = c_3$, for a decision variable $x$ and constant values $c_1, c_2$ and $c_3$. Each FPM rule also has an associated significance and unsignificance where the significance indicates the fraction of solutions in the *selected* set that live up to the rule (the support of the rule in the *selected* set), and the unsignificance indicates the fraction of solutions in the *unselected* set that corresponds to the rule (the support in the *unselected* set). A meaningful and descriptive rule would have a high significance while having a low unsignificance, thereby describing more solutions in the *selected* set. Additionally, using frequent itemset mining, single rules can be combined to find rule-interactions and their significance and unsignifiacnce. Consider the following example of a 3-level rule-interaction: $\{x_1 < 0.2 \wedge x_2 > 0.3 \wedge x_3 = 4\}$ with significance equals to 90% and unsignificance equal to 5%, it says that the combination of these three rules covers 90% of the preferred solutions, while only capturing 5% of the remaining (unpreferred) ones.

## *2.4 Concluding remarks*

To the best of our knowledge, most of the prior studies have neglected the real-world uncertainty and the buffers consideration. Most of the optimization studies adopted meta-heuristics algorithms. However, the use of MOO is sporadic, and the RMS challenging areas are rarely tackled simultaneously. The use of SBO is scarce, usually focuses on small single objective cases, and mostly targets one main area. Nearly all prior work that combined simulation and optimization required a manual data transfer from one to the other (Petroodi et al. 2019). This constitutes an evident research gap in the use of SMO to combine tasks and resource assignment of scalable MPFL-RMS while considering the buffer allocation problem as additional decision variables and the unreliability of the resources.

Knowledge discovery methods have been applied to extract patterns from manufacturing systems. Studies have shown that association rules extracted from applying data mining and knowledge discovery methods to historical datasets can boost the effectiveness of production development and constitutes a challenging area for the future of manufacturing systems (Kou and Xi 2018; Tripathi et al. 2021). Despite RMS constituting one of the critical enablers that impact significantly on today's so-needed changeable manufacturing systems, the knowledge-capturing and decision-making process is very complex due to their intrinsic complexity and stochastic nature (H. ElMaraghy et al. 2021; Algeddawy and Elmaraghy 2011; Diaz, Aslam, and Ng 2021). Accordingly, considering that the optimization of RMS involves a large number of decision variables and setting-up changeable optimization scenarios is a time-consuming task, the applicability of knowledge discovery methods to RMS applications becomes even more crucial and indicates a research opportunity in order to support decision-makers (H. ElMaraghy et al. 2021; Koren, Gu, and Guo 2018a).

Thus, this article aims to mitigate the above gaps by optimizing the tasks and resource assignment of scalable MPFL-RMS while considering the buffer allocation problem by developing a customized SMO approach enhanced with a novel FPM method for conducting post-optimal analyzes.

## 3  A MOO problem formulation for RMS

This study originated from the challenges enterprises face when adjusting the production resources so that MPFL-RMS can efficiently address volumes and capacity changes. Many factors must be considered, including the work task and resource reconfiguration to maximize THP and minimize TBC. As the parts mix and volume change, the RMS evolves accordingly. The problem assumptions are as follows:

- A MPFL-RMS that consists of one or several WSs, manufactures several products under different production volumes.
- Resources within the RMS are subjected to maintenance, breakdown, setup times, and variability of the tasks times.

- All resources within a WS are identical and perform the same sequence of tasks.
- There are reserved places for adding or relocating resources in the WSs.
- There are inter-station buffers with variable capacity.
- Tasks are subjected to precedence relationship and technological requirements that ensure a feasibility sequence to be performed in specific WSs.

The following notations and their definitions are used while dealing with formulating and optimizing the MPFL-RMS.

| Notations | Definition |
|---|---|
| **Indices:** | |
| $i, r$ | task index |
| $j$ | WS index |
| $m$ | resources index |
| $v$ | variant index |
| **Parameters:** | |
| $NS$ | number of WSs |
| $NV$ | number of variants |
| $NT_v$ | number of tasks for variant $v$ ($v = 1, \ldots, NV$) |
| $TNM$ | total number of resources in RMS |
| $NMWS_{max}$ | maximum number of resources per WS |
| $NMWS_{min}$ | minimum number of resources per WS |
| $NB$ | number of buffers ($NS - 1$) |
| $B_{min}$ | minimum safety buffer |
| $B_{max}$ | maximum buffer capacity |
| $Bunit$ | buffer incremental unit |
| $PR_{irv}$ | precedence relationships for variant $v$; 1 if task $i$ is the predecessor of task $r$; otherwise 0 |
| $TR_{jiv}$ | technological requirement for variant $v$; 1 if task $i$ can be assigned to WS $j$; otherwise 0 |
| $THP$ | throughput per hour |
| $TBC$ | total buffer capacity |
| **Decision variables:** | |
| $x_{ijv}$ | 1 if task $i$ is assigned to WS $j$ for variant $v$; 0 otherwise |
| $y_{mj}$ | 1 if resource $m$ is assigned to WS $j$; 0 otherwise |
| $B_j$ | in-between buffer capacity for WS $j$ and $j + 1$ |

The problem formulation is presented below. Two conflicting optimization objectives are defined:

$$Maximize\ f1 = THP = \#products/(simulation\ horizon - warmup) \quad (1)$$

$$Minimize\ f2 = TBC = \sum_{j=2}^{NS} B_{j-1} \quad (2)$$

The following constraints must be satisfied when optimizing the MPFL-RMS.

Task assignment: for each variant $v$, each task must only be assigned to one WS:

$$\sum_{j=1}^{NS} x_{ijv} = 1, \quad \forall v, \forall i = 1,2, \ldots, NT_v \quad (3)$$

Precedence relationships: for each variant $v$, each task can only be assigned to a WS only if all its predecessors are assigned to the same WS or earlier:

$$\sum_{j=1}^{NS} j \times (x_{rjv} - x_{ijv}) \leq 0, \quad \forall v, \forall (r,i) = 1,2,\ldots, NT_v \in \{PR_{irv}|\ PR_{irv} = 1\} \quad (4)$$

Resource assignment: each resource must be assigned to a WS:

$$\sum_{j=1}^{NS} y_{mj} = 1, \quad \forall m = 1,2,\ldots, TNM \quad (5)$$

Technological requirement: for each variant $v$, each task can only be assigned to a WS if it has the required machinery to perform the task:

$$x_{ijv} \leq TR_{jiv}; \quad \forall v = 1,\ldots, NV, \forall i = 1,2,\ldots, NT_v, \forall j = 1,2,\ldots, NS \quad (6)$$

Minimum WS equipment: at least $NMWS_{min}$ should be assigned to each WS:

$$\sum_{m=1}^{TNM} y_{mj} \geq NMWS_{min}, \quad \forall j = 1,2,\ldots, NS \quad (7)$$

Maximum WS equipment: WSs cannot have more than a certain number of resources:

$$\sum_{m=1}^{TNM} y_{mj} \leq NMWS_{max}, \quad \forall j = 1,2,\ldots, NS \quad (8)$$

Buffer capacity: The inter-station buffers should not be less than the minimum safety buffer ($B_{min}$) and not exceed the maximum buffer capacity ($B_{max}$):

$$B_{min} \leq B_{j-1} \leq B_{max} \quad j = 2,\ldots, NS \quad (9)$$

Because the considered MPFL-RMS belongs to NP-hard optimization problems, the next section proposes a SMO approach to address it.

## 4  A simulation-based multi-objective optimization approach

The SMO proposed in this paper consists of two major components, simulation and optimization. The simulation component consists of a DES software named FACTS Analyzer (Ng et al. 2011) in which the RMS and the studied scenarios are modelled and simulated. The optimization component is implemented in the well-known platform MATLAB where the assignment of tasks and resources to WS is performed. The tight integration of the simulation and optimization components allows an accurate representation of a realistic RMS, including many types of system variables regardless of their nature (e.g., failures, mean time to repair, resources' availability, process time, etc.) while avoiding simplifying the RMS as seen in other optimization studies.

The process starts in the optimization component, in which a population of size $NP$ of priority-based RMS solutions is generated. Then, the custom-made RMS-specific encoding and decoding mechanisms are used to generate feasible RMS solutions. Then, the generated population of feasible solutions is mapped to the simulation component in which the RMS configurations are generated based on the received combination of inputs parameters from the optimization. After these solutions are simulated, the results of the simulation-based fitness function evaluation in term of multiple objectives are fed back to the optimization component. Next, by using a random solutions selection mechanism, the population of solutions goes through the crossover and the mutation operators based on specific probabilities $c_p$ and $m_p$, respectively, to generate a new population of offspring. The above iterative process is repeated until the integrated optimization and simulation components converge to a set of Pareto-optimal solutions or the stopping criteria, a prespecified number of generations ($G_{max}$), is reached. The main structure of the proposed SMO approach is illustrated in Figure 1.

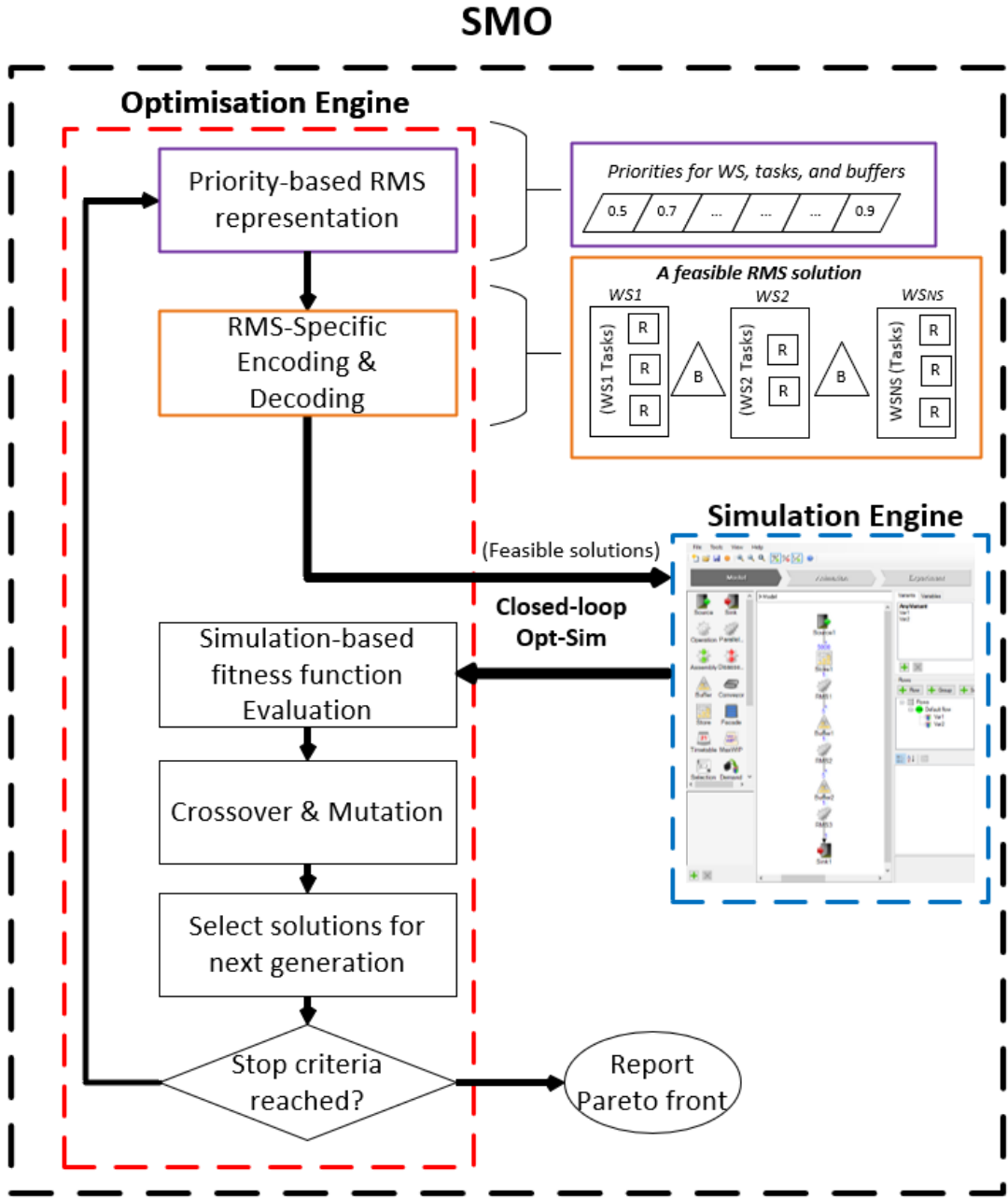

Figure 1. SMO approach.

In the optimization context, the metaheuristic algorithms have been proven promising approaches for any combinatorial optimization problem. Among metaheuristics, GAs have been extensively employed to optimize manufacturing systems (Lidberg et al. 2020). When dealing with two conflicting objectives, the NSGA-II is known to be one of the most effective MOEA endowing a proper convergence and spread of solutions (Michalos, Makris, and Mourtzis 2012). Three main factors drive the outstanding performance of NSGA-II; the fast non-dominated sorting approach that decreases the computational complexity; the elitism mechanism storing the non-dominated solutions; and the crowding distance calculation that ensures a diverse population by comparing and selecting solutions after the non-dominated sorting (Ng, Bernedixen, and Syberfeldt 2012). In this study, by incorporating the simulation components into the fitness function evaluation of NSGA-II, a customized SMO-NSGA-II for RMS is developed, as described below.

### *4.1 SMO-NSGA-II for RMS*

The NSGA-II sorts the solutions into different fronts based on their dominance relationship (dominated and non-dominated). The dominance relationship is established between each pair of solutions by comparing the objectives set by the objective functions. The crowding distance ensures a good spread of solutions by determining the density in the region, impacting the selection of the solutions that will be preserved for future generations. Once the fast non-dominated sorting is completed, the crowding distance calculation ranks the solutions in each individual front. The above features of NSGA-II promote the selection of dispersed solutions on the fronts. The proposed SMO-NSGA-II for RMS is summarized in Pseudo-code 1.

| Pseudo-code 1: SMO-NSGA-II | |
|---|---|
| 1 | Algorithm inputs: $G_{max}$, $NP$, $c_p$, $m_p$ <br> RMS inputs: $TR_{j,iv}$, $PR_{irv}$, $NS$, $NT_v$, $TNM$, $NMWS_{min}$, $NMWS_{maxn}$, $NB$, $B_{min}$, $B_{max}$, $Bunit$ |
| 2 | Using Section 4.1.1, create a population of priority-based representation vectors |
| 3 | While $g \leq G_{max}$ |
| 4 | Using the proposed encoding and decoding mechanisms in Section 4.1.2 to ensure a population of RMS feasible solutions |
| 5 | Using the simulation component in Section, 4.1.3, evaluate the fitness function for each solution |
| 6 | Rank the solutions using the fast non-dominated sorting mechanism |
| 7 | Calculate the crowding distance of each solution in each individual front |
| 8 | Select parents for crossover using tournament selection |
| 9 | Using crossover and mutation operators in Section 4.1.4, generate a new set of offspring |
| 10 | Using the elitism mechanism to preserve the best individuals |
| 11 | Increment $g$ |
| 12 | End |
| 13 | **Output**: The Pareto-optimal solutions for RMS |

The components of the SMO-NSGA-II for RMS are described below.

### *4.1.1 Solution representation:*

The NSGA-II starts with an initial population of the individual solutions in which each row represents a string of real numbers ($\sigma$) where the elements are randomly generated between (0,1). The length of the solution strings is equal to the number of WSs ($NS$) plus the number of tasks for all the variants ($\sum_{v \in NV} NT_v$) plus the number of inter-station buffers ($NB=NS-1$). The bit content at $ind$-th index called $\sigma_{ind}$ ($ind=1,...,NS+\sum_{v \in NV} NT_v+NB$), contains the random number showing the relative priority of WSs, tasks, and buffers depending on where the $i$ index relies, as depicted in Figure 2. As the figure shows, the first NS columns relate to the WSs priority meaning that the higher the priority, the more resources will be assigned to the WS. The same priority rule applies to the buffers in the last *NB* columns of the string. The random-keys from column $NS+1$ to $NS+\sum_{v \in NV} NT_v$ relate to the priority of the tasks, indicating that a task with a higher relative priority value is ranked higher to be assigned to the WSs. Figure 2 illustrates the representation for an example solution with two WSs, one in-inter-station buffer, and two parts to be produced with 2 and 3 tasks, respectively.

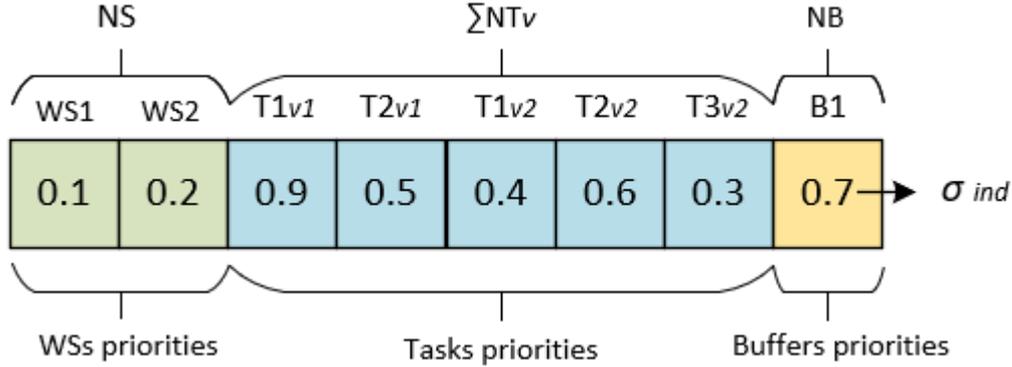

Figure 2. Solution representation.

### 4.1.2 Encoding and Decoding:

The encoding and decoding procedures aim at generating a feasible solution for each solution string in the population.

For each string, the encoding attempts to find feasible settings for the RMS configuration by assigning resources to the WSs, deciding on the assignment order of the tasks to WSs, and assigning inter-station buffer capacities. The number of resources per WS is calculated using $\lceil NMWS_{min} + \sigma_{ind} \times (NMWS_{maxn} - NMWS_{min}) \rceil$ where $\sigma_{ind}$ $(ind = 1, \dots, NS)$ refers to the WSs priorities and $\lceil \ \rceil$ is the lowest bigger integer number. If the total number of resources per WS is not equal to the $TNM$, they are updated until their sum equals $TNM$. The assignment order of the tasks to WSs for each variant $v$, is decided by the flexibilities of the tasks, based on $TR_{jiv}$, and the proprieties of the tasks, based on $\sigma_{ind}$ $(ind = NS + 1, \dots, NS + \sum_{v \in NV} NT_v)$, in ascending and descending orders, respectively. The in-station buffer capacity is calculated using $\lceil B_{min} + \sigma_{ind} \times (B_{max} - B_{min}) \rceil$ where $\sigma_{ind}$ $(ind = NS + \sum_{v \in NV} NT_v + 1, \dots, NS + \sum_{v \in NV} NT_v + NB)$ refers to the priorities of the buffers. If the summation of the inter-station buffer capacities is less than $B_{min} \times NB$ or larger than $B_{max} \times NB$, then they are updated until they fall in the range above. The encoding procedure is shown in Pseudo-code 2.

```
Pseudo-code 2: Encoding
1   Input: σ, TR_{jiv}, NS, NT_v, TNM, NMWS_{min}, NMWS_{maxn}, NB, B_{min}, B_{max}, Bunit
2   For ind = 1 to end
3       If ind = 1 to NS
4           Calculate the number of assigned resources per WS based on σ_{ind}, NMWS_{min}, and
            NMWS_{maxn}
5           If the total number of assigned resources > TNM
6               Sort WSs in terms of their σ_{ind} in descending order
7               While the total number of assigned resources > TNM
8                   Decrease one resource from the sorted WSs in line 6
9               End
10          Elseif the total number of assigned resources < TNM
11              Sort WSs in terms of their σ_{ind} in ascending order
12              While the total number of assigned resources < TNM
13                  Increase one resource to the sorted WSs in line 11
14              End
15          End
16      Elseif ind = NS + 1 to NS + Σ_{v∈NV} NT_v
17          For v = 1 to NV
18              Sort the tasks of variant v in terms of their flexibility (based on TR_{jiv}) and priority
                (based on σ_{ind}) in ascending and descending orders, respectively
19          End
20      Elseif ind = NS + Σ_{v∈NV} NT_v + 1 to end
21          Calculate the allocated in-between WSs buffer capacity based on σ_{ind}, NB, B_{min}, and
            B_{max}
22          If the total allocated buffers capacity > B_{max} × NB
23              Sort in-between buffers in terms of their σ_{ind} in descending order
24              While the total number of in-between buffers capacity > B_{max} × NB
25                  Decrease one Bunit from the sorted in-between buffers in line 23
26              End
27          Elseif the total allocated buffers capacity < B_{max} × NB
28              Sort in-between buffers in terms of their σ_{ind} in ascending order
29              While the total number of in-between buffers capacity < B_{min} × NB
30                  Increase one Bunit to the sorted in-between buffers in line 28
31              End
32          End
33      End
34  End
35  Output: number of resources per WS, vectors of sorted tasks based on flexibility and priority
    per variant, in-between buffers capacity
```

For each setting obtained via the encoding procedure, the decoding aims to generate a feasible solution for the RMS by assigning tasks to WSs, considering the vectors of the sorted tasks based on the flexibility and the priority for each variant. This is performed by selecting each task of variant $v$ based on the related vector of sorted tasks and then positioning the task within the eligible range of WSs considering the cumulative normalized vector of $TR_{jiv}$ and their priorities $\sigma$. The decoding procedure is shown in Pseudo-code 3.

```
Pseudo-code 3: Decoding
1    Input: PR_{i rv}, TR_{jiv}, NS, NT_v, vectors of sorted tasks based on flexibility and priority per
     variant
2    For v = 1 to NV
3        For i = 1 to NT_v
4            task= set the selected task as the ith index in the sorted tasks vector for variant v
5            Normalize the TR_{jiv} matrix by TR2_{jiv} = TR_{jiv}/∑_{j∈NS} TR_{jiv}
6            Calculate the accumulative sum for TR2_{jiv} by TR3_{jiv} = ∑_{j=1}^{j2} TR2_{jiv} where j2 = 1 to
             NS
7            Station=min_j find{TR3_{j task v} ≥ σ_{task}}; set the selected station by finding the next
             index where TR3_{j task v} is bigger than σ_{task} for variant v
8            Update TR_{jiv} after the current task has been assigned to WS for variant v including all
             its predecessors and successors in PR_{i rv}
9        End
10   End
11   Calculate the total task time per variant per WS
     Output: A feasible solution for RMS includes the number of resources per WS (encoding),
12   assignment of tasks to WSs (decoding), the total task time per variant per WS (decoding), and
     in-between buffers capacity (encoding)
```

*4.1.3  SMO-based fitness function evaluation of RMS solution*

To guide the optimization and enable NSGA-II to perform the non-dominated ranking, the simulation measures and provides the fitness function of the solutions. To this end, each RMS solution is mapped to the simulation component, where a simulation scenario is built according to the information received from the optimization. These sets of scenarios are simulated, including the production variabilities in terms of availability of resources, failures, setup times, and production proportions. For each scenario, the simulation component calculates the objectives function values in terms of THP and TBC before they are fetched back to the optimization component.

*4.1.4  Genetic operators (crossover/mutation)*

To preserve the diversity between generations, the genetic operator takes place randomly in each generation, as inspired by the biological processes. To ensure that the best solutions are more likely to get more copies, the best solutions are selected using the Tournament selection to be preserved based on the SMO-based fitness function evaluation. Then, they go through the crossover and the mutation operators to generate a new diverse population. The crossover and mutation probabilities ($c_p$ and $m_p$) of the genetic operators indicates the percentage of the population that will go through these processes respectively.

    A two-points-based weight mapping crossover is implemented. This crossover operator can be explained in four steps. The first step chooses two intervals randomly on the chromosomes of two selected solutions (parents). In the second step, the bits included in the crossover interval are ranked in ascending order based on their priority values. A lower ranking value indicates a bit with a higher priority. In the third step, the ranks between the chosen intervals are swapped between the parents, and the priorities are rearranged based on the new ranks. Therefore, the offspring are generated according to the newly mapped priorities in the four steps. The upper part of Figure 3 illustrates the implemented crossover steps.

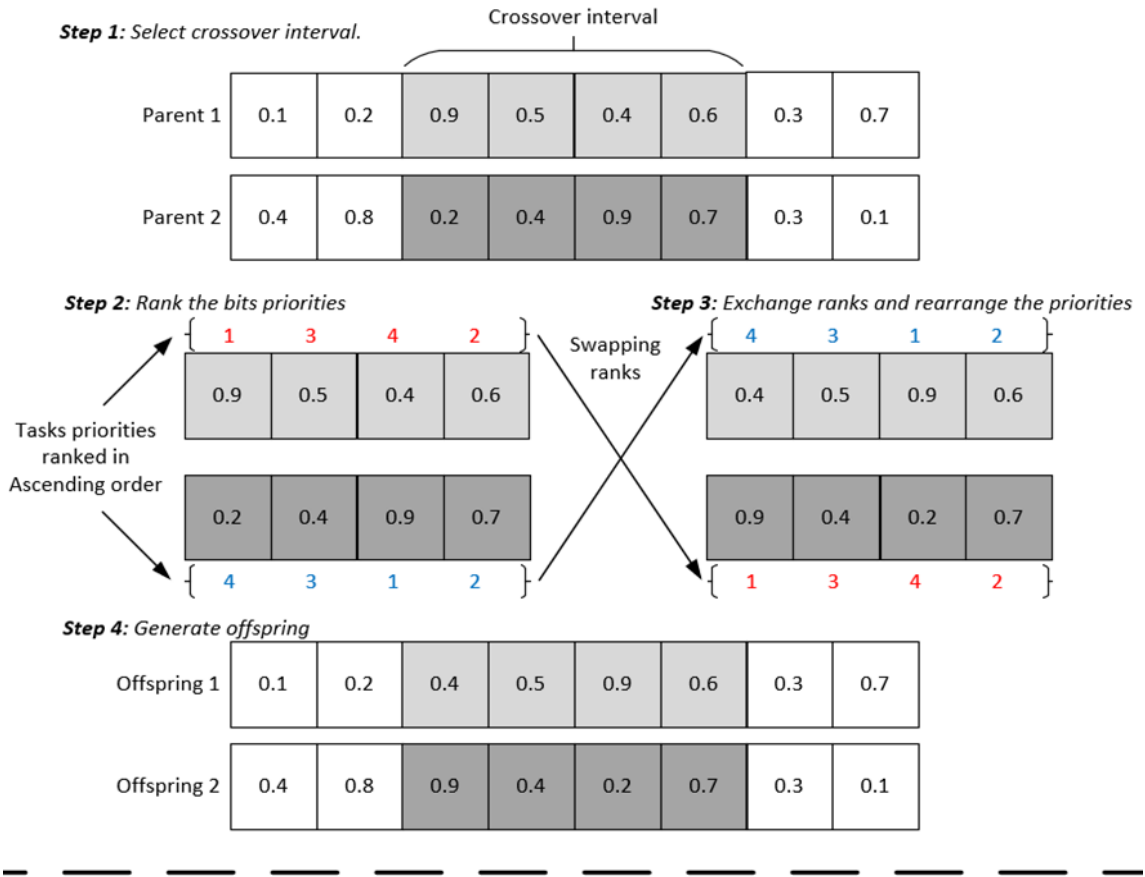

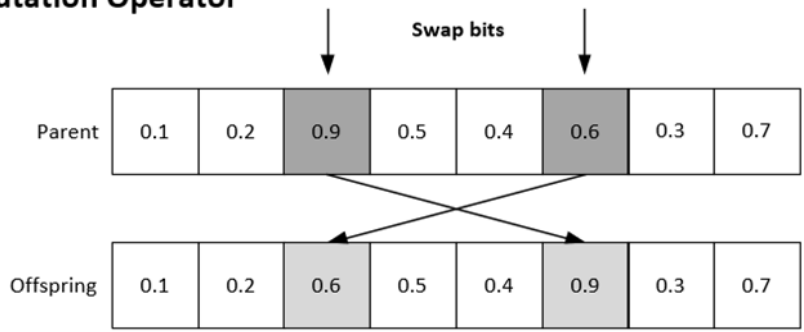

Figure 3. Genetic operators.

The implemented mutation operator swaps the values of two randomly selected bits. The not selected bits are preserved from the parent. The bottom part of Figure 3 illustrates the mutation operator, where the darker bits represent those that are swapped.

The applicability of the proposed SMO-NSGA-II approach is shown using an application case described in the next section.

## 5   An application case

The case is based on a MPFL at a R&D facility of a truck manufacturer in Sweden, where the manufacturer tests and evaluates future concepts. The system manufactures two product families. The company has invested in 3 reconfigurable WSs in which the resources can be added, removed, or reallocated to a different WS if required due to production changes. Each WS has space for up to 5 resources (e.g., operators). Figure 4 represents an example of the MPFL when they have 7 operators configured in a 3-2-2 setting, meaning three operators in the first WS, and two in each of the remaining WSs. Notice that there are remaining extra spaces for resources 2, 3, and 3 in the first, second, and third WS, respectively. Furthermore, the WSs of this MPFL are subjected to uncertainty and variability, and they consider a specific availability and mean time to repair (MTTR). The availability is considered to be 85 % with a 10-minute MTTR. There are two inter-station buffers with a minimum safety capacity of $B_{min}=1$ and a maximum buffer capacity of $B_{max} = 40$. The buffers' system admits capacity increments in the steps of one product. Additionally, the buffers required 5 seconds for loading/unloading, as the material handling times.

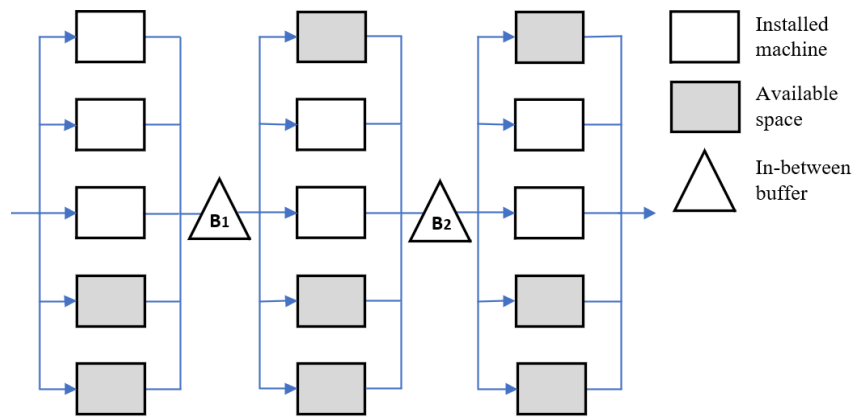

Figure 4. MPFL layout.

In the studied case, the two product families must be produced at specific volumes to meet the customers' demands. As the customers' demands vary over time, the MPFL configuration, the components of the system, and the process plan evolve accordingly to meet the new requirements. The changes in the line involve not just the total number of resources needed and the layout configuration in the WSs, but also the assignment of the tasks to them. Moreover, to accommodate a new demand requirement, a configuration change is needed, which also implies changes in buffer capacity. The manufacturing company was interested in finding out the production capacity of the MPFL with an initial investment of 7 operators for the production of different proportions, 70/30 (70% part 1 and 30% part 2) and 30/70. In addition, the company also requested information regarding the capacity that could be gained if 1 and 2 operators were added to the system, including where they need to be placed according to the desired proportion and the new optimized tasks assignment of both parts. This study strives for maximum THP and minimum TBC as the optimization objectives simultaneously while deciding on the capacity of the buffers in the line and the allocation of resources and tasks to WSs according to the desired scenarios.

The total task time of part 1 is 336.38 seconds, divided into 29 tasks while the total processing time of part 2 is 293.38 seconds, divided into 24 tasks. Figure 5 shows the precedence relationship of the tasks for both parts.

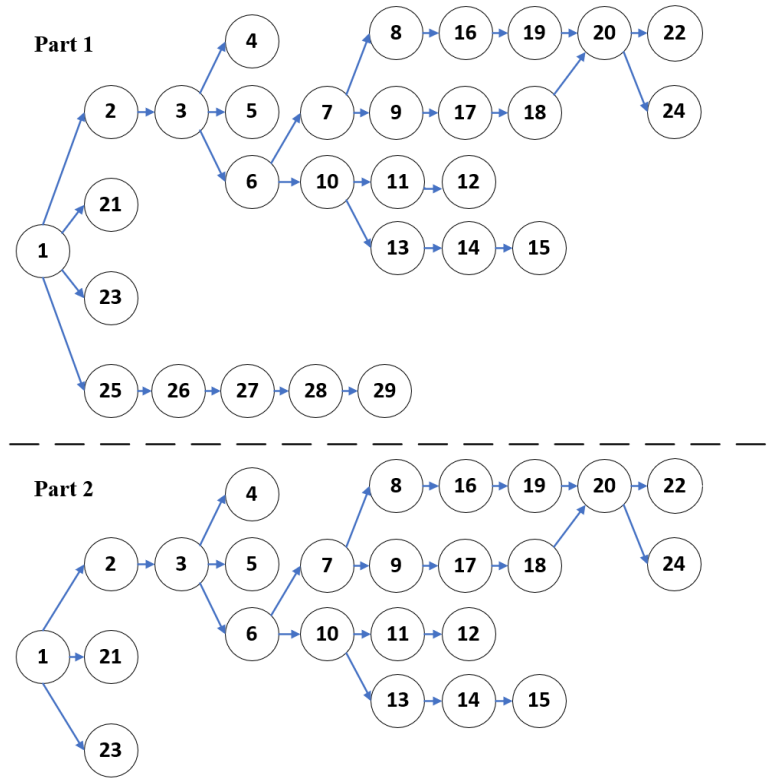

Figure 5. Precedence graphs.

## 6 Experimental results and knowledge discovered

The SMO-NSGA-II approach is implemented in MATLAB VERSION 2022a and Facts Analyzer version 3.1.7. The experiments include six different scenarios that investigate the above-explained RMS under variable production proportions utilizing a scalable number of operators. A baseline simulation model of the RMS was developed in the mentioned DES software to be used in the proposed SMO approach. Every scenario was optimized using 500 generations and a population size of 50.

Figure 6 illustrates the objective space of the non-dominated solutions found by the proposed SMO-NSGA-II approach for RMS regarding the studied scenarios. According to this figure and as it was expected, the more operators used, the higher the THP of the system.

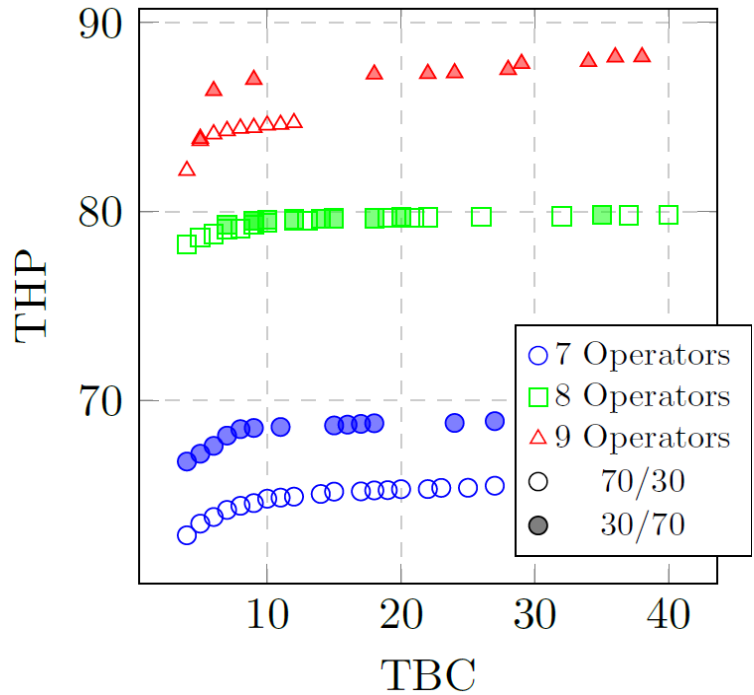

Figure 6. Objective space of non-dominated solutions.

For a better explanation of the results in Figure 6, the obtained ranges for the THP, the capacity of the inter-station buffers ($Bu_1$ and $Bu_2$), and the TBC (sum of $Bu_1$ and $Bu_2$) for each scenario, are shown in Table 1. Each scenario in the table is characterized by the number of operators used (NO) and the production proportion. Considering the maximum THP obtained as the results of different scenarios shown in Table 1, one can observe that the optimized average THP increase that can be gained from every operator added to the considered RMS, is approximately 9.64 JPH (Jobs Per Hour) for 30/70 proportion and 8.11 JPH for 70/30 proportion respectively. This is important for the engineers to consider when scaling up (or down) the system to adjust the production volume required.

Table 1. Throughput and buffers capacity ranges.

| NO | Proportion | THP | $Bu_1$ | $Bu_2$ | TBC |
|---|---|---|---|---|---|
| 7 | 30/70 | 66.76-68.90 | 2-13 | 2-14 | 4-27 |
|   | 70/30 | 62.86-65.48 | 2-6 | 2-21 | 4-27 |
| 8 | 30/70 | 79.31-79.82 | 3-7 | 4-28 | 7-35 |
|   | 70/30 | 78.27-79.81 | 2-16 | 2-24 | 4-40 |
| 9 | 30/70 | 83.86-88.18 | 2-5 | 3-33 | 5-38 |
|   | 70/30 | 82.16-84.69 | 2-5 | 2-7 | 4-12 |

Table 2 presents how the results presented in Table 1 are attained in terms of the system configuration (Operators per WS) and the tasks allocation per WS. Under columns WS1, WS2, and WS3, the number of parallel operators employed in workstations 1, 2, and 3 are presented, respectively. The last column shows the number of tasks performed in each WS (i.e., no. tasks allocated to WS1/ no. tasks allocated to WS3/ no. tasks allocated to WS3). Note that the more operators in a WS, the more tasks are assigned to it. Additionally, it is shown that in most cases, the number of tasks per WS ranges among the non-dominated solutions.

Table 2. Configuration and work task allocation.

| NO | Proportion | WS1 | WS2 | WS3 | Tasks |
|---|---|---|---|---|---|
| 7 | 30/70 | 1 | 2 | 4 | 9/14-15/29-30 |
|   | 70/30 | 1 | 2 | 4 | 8/11-12/33-34 |
| 8 | 30/70 | 2 | 3 | 3 | 13-16/17-20/19-20 |
|   | 70/30 | 3 | 2 | 3 | 19-20/12-14/20-21 |
| 9 | 30/70 | 2 | 3 | 4 | 13-14/12-14/26-28 |
|   | 70/30 | 2 | 3 | 4 | 15-16/13-15/23-24 |

An overview of a core tenant of this approach, the assignment of tasks to WSs and the related pattern in the non-dominated solutions, is presented in Figure 7. In this figure, each row represents one solution, and each column illustrates one task for either part 1 or 2. Moreover, the color of the cells indicates the WS where the related task (A indicate tasks from part 1 and E tasks from part 2) of each part has been assigned. The figure shows how most non-dominated solutions for each scenario share a common task allocation. Nonetheless, all solutions shown in Figure 7 are distinct in terms of the allocation of the inter-station buffers.

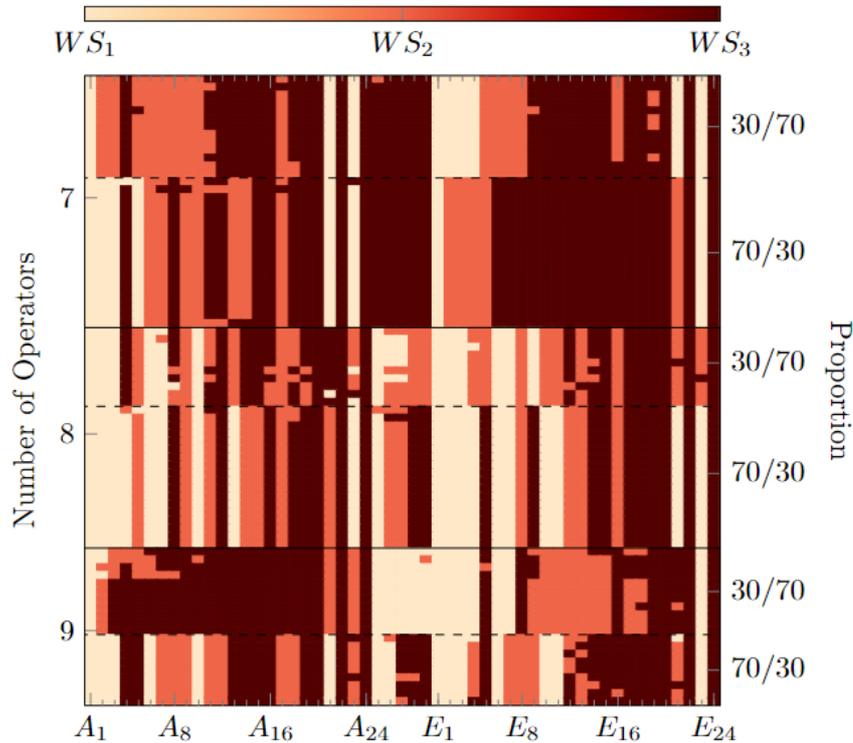

Figure 7. Task assignment in the non-dominated solutions RMS scenarios.

## 6.1 Approach comparison

In this subsection, the proposed SMO-NSGA-II approach is compared with a standard

SMO approach presented by (Diaz, Aslam, and Ng 2021) for RMS, in which both optimization and simulation are run on standard SMO. This comparison aims to explore whether the proposed SMO with customized procedures improve the resulting RMS solutions. To this end, the same scenarios were implemented and optimized by using the standard SMO while using the same algorithm settings for the considered RMS application. The abovementioned study proved the standard SMO to be effective for an industrial application RMS. However, the total number of decision variables was much lower mainly due to the total number of tasks of the produced products being 25 compared to the 53 involved in the current RMS. The standard SMO uses a commercial NSGA-II embedded in the software that does not allow the customization of specific procedures such as encoding and decoding which leads to the use of a repair mechanism.

The comparison of the convergence rates of the standard SMO and the proposed SMO approaches is shown in Figure 8 by plotting the Hypervolume (HV) (Zitzler and Thiele 1999) of the optimization algorithms at each generation. The higher the HV, the better the quality of the obtained solutions. According to this figure, one can evidently observe that the proposed SMO has a better convergence than the standard SMO in finding Pareto-optimal RMS solutions.

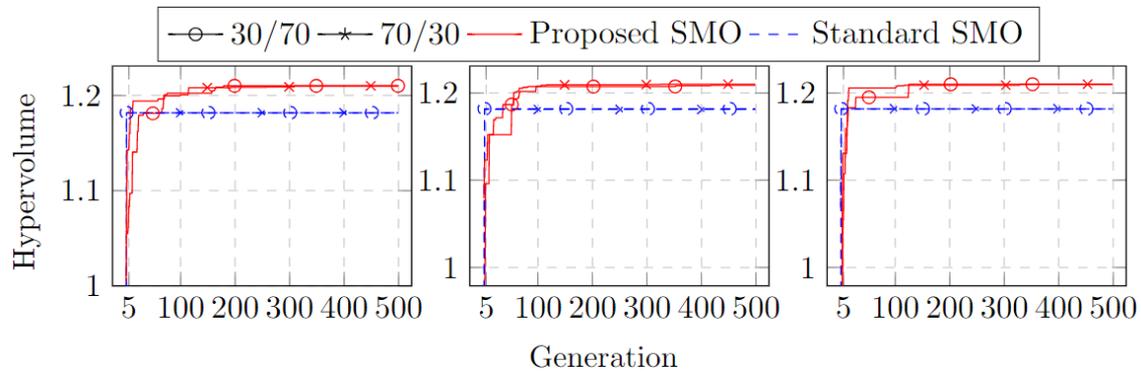

Figure 8. Convergence rate plots of standard and proposed SMO for the scenarios with 7 operators (left-hand side), 8 operators (center), and 9 operators (right-hand side).

Furthermore, the HV of the non-dominated solutions obtained by the optimization approaches for different scenarios, including the number of operators (NO) and the production proportions, are shown in Table 3. The comparison of results in Table 3 indicates that in all considered scenarios, a considerable improvement in HV has been achieved when the proposed SMO was applied. These improvements can be explained by the customization of NSGA-II performed in the proposed SMO, enabling the optimization algorithm to deal with larger and more complex RMS applications.

Table 3. Quantitative HV comparison.

| NO | Proportion | Proposed SMO | Standard SMO |
|---|---|---|---|
| 7 | 30/70 | **1.066E+00** | 3.277E-01 |
|   | 70/30 | **1.055E+00** | 3.516E-01 |
| 8 | 30/70 | **1.031E+00** | 2.981E-01 |
|   | 70/30 | **1.159E+00** | 1.927E-01 |
| 9 | 30/70 | **1.070E+00** | 3.512E-01 |
|   | 70/30 | **9.341E-01** | 3.072E-01 |

## 6.2 Knowledge Discovery

This subsection presents the knowledge discovered by applying FPM to datasets generated by the proposed optimization approach. Due to the ability of RMS to increase and decrease the number of resources to address, among other challenges, fluctuating production volumes, this study focused on discovering generalized knowledge regarding the different number of operators employed and the production proportions for the two products in the considered RMS application. Therefore, to run the FPM procedure, the scenarios were merged into five groups in terms of the different number of operators and production proportions. FPM generates knowledge in the form of decision-rules, and we focused on generating knowledge for both the task allocation and the WS and buffers allocation.

For each group, we run FPM with the non-dominated solutions from the involved scenarios as the selected set while keeping the remaining solutions (dominated and non-dominated) as the unselected set. In this way, the general knowledge between the scenarios is discovered. Because of the high number of decision variables involved, which impacts the run-time of the FPM procedure, the maximum level of rule interactions was limited to 5, and the minimum required significance of the rules that describe the selected set was set to 90%. The openly available decision support tool Mimer[1] was employed to generate the results. Mimer enables the interactive knowledge discovery framework for MOO proposed in (Smedberg and Bandaru 2022). The rule interactions found by FPM regarding the task allocation are presented in Table 4, where "A" and "E" refers to the related tasks for part 1 and part 2, respectively. The value of the variables represents in which WS the task was assigned. As an example, the first rule of the table for the cases in which seven operators were used states that in 100 % of the solutions found in the Pareto-front, for part 1, task 10 was set to WS 2, while for part 2, task 5 was set to WS 2, task 23 to was set to WS 1, task 4 was not set to WS 3, and task 6 was not set to WS 1.

Table 4. Resulting rules regarding the work task allocation.

| Scenario | | Rule-interaction | Sig. | Unsig. |
|---|---|---|---|---|
| NO | Proportion | | | |
| 7 | | $A_{10} = 2 \land E_4 \neq 3 \land E_5 = 2 \land E_6 \neq 1 \land E_{23} = 1$ | 100% | 10.49% |
| 8 | | $A_{10} = 1 \land A_{14} \neq 1 \land A_{17} = 2 \land E_7 = 1 \land E_{16} = 2$ | 100% | 15.44% |
| 9 | | $A_{23} = 2 \land A_{26} = 1 \land E_6 \neq 1 \land E_9 \neq 1 \land E_{23} = 1$ | 90.00% | 11.91% |
| | 30/70 | $A_2 \neq 3 \land A_{14} = 3 \land E_3 = 1 \land E_{10} \neq 1 \land E_{23} \neq 3$ | 97.06% | 29.98% |
| | 70/30 | $A_3 = 1 \land E_3 \neq 3 \land E_{11} \neq 2 \land E_{13} \neq 1 \land E_{23} = 1$ | 100% | 23.97% |

Looking at the rules presented in Table 4, we can see that for the scenarios with 7 operators having the highest ratio significance of 100% and unselected significance of 10.49%, meaning that all non-dominated solutions support the rules found while only 10.49% of the unselected set of solutions support the rule-interaction. This implies that the non-dominated solutions in these scenarios are perhaps easier to distinguish than the rest. Another spotted aspect from the rule of the seven operators' scenarios is a higher involvement of tasks from part 2 than from part 1. This could indicate that part 2 needs to be prioritized over part 1 when seven operators are employed regardless of the produced proportion. Additionally, it can also be seen that the scenarios focused on the number of operators have a lower unselected significance than those focused on the proportions. Therefore, more general knowledge is extracted regarding the number of operators employed in the system than the proportion produced.

---
[1] https://assar.his.se/mimer/html/

Decision-makers can use the results presented in the table to know which tasks to prioritize when a new scenario needs to be optimized. Another interesting aspect extracted is the importance of some tasks. Looking at the table, it can be seen that task $E_{23}$ (task 23 of part 2) is repeated in almost all the rules suggesting the relevance of this task for the overall RMS. Furthermore, as can be interpreted from the rules, $E_{23}$ does not take the value 3; in fact, in all the cases, but one is equal to 1, meaning that this specific task should be allocated rather at the beginning of the RMS and never in the last WS. Furthermore, Table 4 shows that for the found rules of all studied scenarios, except for the scenario with eight operators, there is a higher involvement of part 2 than part 1. This suggests that decision-makers could prioritize the assignment of the tasks of part 2 over part 1, even in cases where the production of part 1 is bigger such as 70% of part 1 and 30% of part 2 production volumes. Consequently, with this information, decision-makers can better understand which decision variables are more critical and how they impact the overall performance of the system. It is important to note that the rules in the table best distinguish the selected set from the unselected set. A rule would not be interesting if it has both a high significance and a high unselected significance, only the rules unique to the non-dominated solutions in each group of scenarios distinguish the selected and unselected sets.

Table 5 presents the discovered rules regarding WSs and buffer allocation. Similar to the rules describing the task allocation presented in table 4, the unselected significance of the results in table 5 confirms that it is more difficult to generalize knowledge regarding the proportion compared to the number of operators used. Additionally, we can see that the unselected significance for the rules in Table 5 is higher than in Table 4, meaning that it is more difficult to distinguish the scenarios based on the workstation and buffer allocation. This is, however, expected since the number of variables considered in Table 4 is much greater (53) than the number of variables considered in Table 5 (5). Furthermore, the rules presented in Table 5 provide information regarding operators' load per WS, hinting at which WSs need a higher or lower number of operators. Likewise, interesting aspects can be extracted from the rules regarding the inter-station buffers. As an example, the rules display information regarding the maximum capacity of each buffer which can be used by decision-makers when deciding on new scenarios. This analysis shows the FPM procedure can provide decision-makers with a better knowledge of the system and consequently save time and cost.

Table 5. Resulting rules regarding the workstations and buffer allocation.

| Scenario | | Rule-interaction | Sig. | Unsig. |
|---|---|---|---|---|
| NO | Proportion | | | |
| 7 | | $WS_1 = 1 \land WS_2 = 2 \land Bu_1 < 7 \land Bu_2 < 22$ | 90.63% | 20.51% |
| 8 | | $WS_3 = 3 \land WS_1 \neq 1 \land Bu_1 < 17 \land Bu_1 > 2$ | 92.86% | 26.15% |
| 9 | | $WS_2 = 3 \land WS_3 = 4 \land Bu_1 < 7 \land Bu_2 < 34$ | 100% | 24.80% |
| | 30/70 | $WS_1 < 3 \land WS_2 \neq 1 \land Bu_1 < 15 \land Bu_2 > 2$ | 94.12% | 71.01% |
| | 70/30 | $WS_2 \neq 1 \land WS_3 > 2 \land Bu_1 < 17 \land Bu_2 < 25$ | 100% | 83.81% |

The employed FPM methods and knowledge discovery can also be used to compare the performance of the SMO approaches. The significance of the rules indicates their quality in achieving better results. For comparison purposes, we run FPM, regarding the work task allocation, to the datasets generated by the standard SMO. Table 6: Resulting rules of the Standard SMO approach regarding the work task allocation. displays the resulting rules. The results presented in this table are particularly different from the rules presented in Table 4, implying that the standard SMO did not converge to optimal solutions.

Table 6: Resulting rules of the Standard SMO approach regarding the work task allocation.

| Scenario NO | Proportion | Rule-interaction | Sig. | Unsig. |
|---|---|---|---|---|
| 7 | | $A_{16} \neq 1 \wedge A_{18} = 3 \wedge E_5 \neq 1 \wedge E_{19} \neq 2 \wedge E_{21} = 2$ | 92.86% | 16.48% |
| 8 | | $A_5 \neq 2 \wedge A_{11} = 2 \wedge A_{13} \neq 2 \wedge A_{28} \neq 2 \wedge E_{19} \neq 2$ | 90.48% | 23.00% |
| 9 | | $A_4 = 3 \wedge A_{12} = 2 \wedge A_{18} \neq 1 \wedge E_4 \neq 2 \wedge E_{19} \neq 3$ | 100% | 6.83% |
| | 30/70 | $A_5 \neq 3 \wedge E_4 \neq 3 \wedge E_5 \neq 1 \wedge E_{11} \neq 2 \wedge E_{16} \neq 1$ | 96.00% | 15.57% |
| | 70/30 | $A_{13} \neq 2 \wedge A_{16} = 2 \wedge A_{21} \neq 1 \wedge A_{28} \neq 1 \wedge E_{18} \neq 1$ | 91.30% | 29.21% |

In addition, the significance/unsignificance relationship of the rules displayed in Table 6 indicates they are less relevant to achieve better results when compared to those displayed in Table 4. This is further exposed for the seven machines scenario in Figure 9. This figure presents both datasets where the circle indicates solutions from the proposed SMO and the square a solution from the standard SMO. The solutions that match the rules for seven machines extracted through the proposed SMO are highlighted in blue, and the solutions that match the rules extracted from the standard SMO are highlighted in red in both datasets. On the one hand, the number of solutions that match the rules extracted from the proposed SMO is significantly smaller than those matching the rules from the standard SMO. Since almost no blue points were found in the standard SMO dataset, this demonstrates the uniqueness and quality of the rules extracted from the proposed SMO approach dataset. On the other hand, many solutions match the rules extracted from the standard SMO regardless of the dataset. This proves that the rules extracted from the standard SMO dataset are not unique and do not represent the optimal solutions. Many of the solutions in the proposed SMO dataset match the rules extracted from the standard SMO; however, as shown in Figure 9, they do not provide the best performance. Consequently, evaluating the quality of the rules extracted using FPM supports the statement from 6.1, where the proposed SMO outperforms the standard SMO.

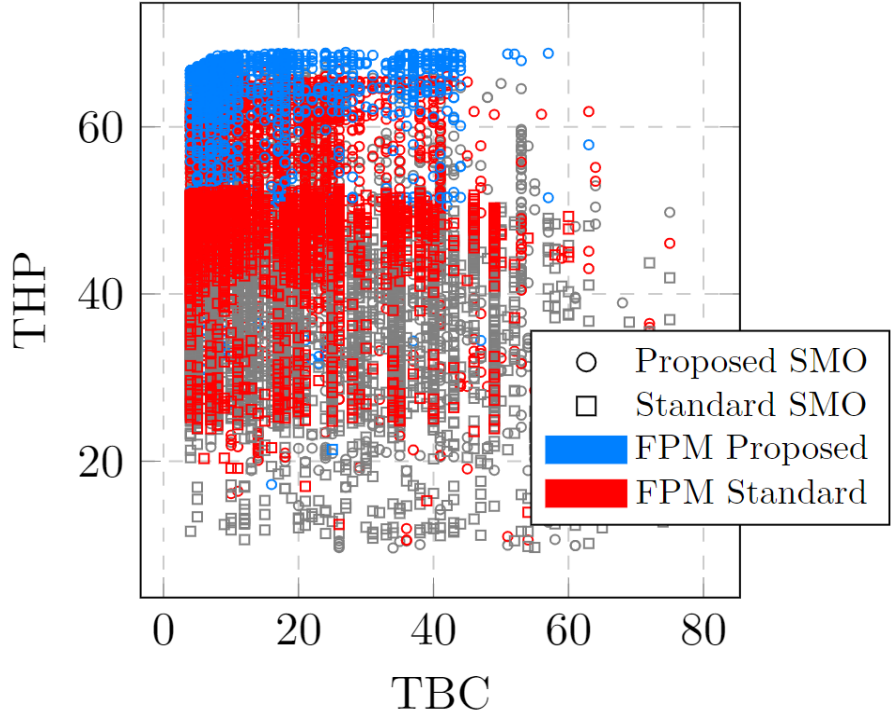

Figure 9: Rules quality comparison for 7 machines scenario.

## 7  Conclusions

In the current uncertainty and competitiveness of market, RMS applications play a significant role in the success of the manufacturing industry. However, prior SMO research that considers the variability of RMS applications is scarce and neglects knowledge discovery to support decision-makers. This study introduced a novel SMO approach to concurrently address the main challenging areas while combining the tasks and resource assignment with the configuration of the system in a scalable MPFL while considering the buffer allocation dilemma as additional decision variables and the unreliability of the system. A customized SMO-NSGA-II approach was developed with problem-specific solution representation, encoding, and decoding mechanisms combined with a simulation component where the RMS solutions are evaluated in terms of multi-objectives, namely, the THP and TBC. The performance of the proposed SMO was tested against standard SMO in the studied application. The experimental results showed the proposed SMO is promising in finding Pareto-optimal solutions compared to the standard SMO. Furthermore, due to the ever-increasing amount of data generated by the MOO of RMS, which is required to address frequent market changes, this study demonstrated how knowledge discovery and data mining methods can be used for extracting rules from RMS applications.

Essentially, the proposed enhanced SMO approach provides fast decision support for RMS production planning, especially when facing fluctuating production volumes. To this extent, the proposed approach support decision-makers with the key information to enable the RMS with the capability to provide the required production capacity when needed. In a nutshell, this approach reveals underlying information that facilitates understanding the RMS and how the decision variable affects the performance of the system. This study used the FPM procedure to generate significant knowledge in the form of decision-rules that describe the task and resource allocation to workstations and buffer capacity allocation for all considered scenarios.

Future research will use the generated knowledge to speed-up the optimization of additional scenarios in a process known as Knowledge-Driven Optimization (KDO) (Bandaru, Ng, and Deb 2017a). Since FPM generates knowledge in the form of explicit decision rules, it would be straightforward for an optimizer to incorporate rules describing a general scenario into a future optimization run by applying the rules as constraints in the decision space. Future work may also consider additional RMS aspects, such as sustainability and reconfiguration frequency.

**Data Availability Statement**

The data that support the findings of this study is available upon request.

**Acknowledgment**

The authors thank the Knowledge Foundation (KKS), Sweden, for funding this research through the KKS Profile Virtual Factories with Knowledge-Driven Optimization, VF-KDO, grant number 2018-0011.

**Disclosure statement**

The authors report there are no competing interests to declare.